\documentclass[fleqn,usenatbib]{mnras}

\usepackage[english]{babel}
\usepackage{amsmath}
\usepackage{amssymb}
\usepackage{graphicx}
\usepackage{physics}
\usepackage{subcaption}
\usepackage{graphicx}
\usepackage{xcolor}
\usepackage{verbatim}
\usepackage{lmodern}
\usepackage{mathtools}

\newcommand{\ann}{\color{red}}

\newcommand{\alfLAW}{\alpha_{\rm law}}
\newcommand{\alfROT}{\alpha_{\rm rot}}
\newcommand{\alfSEG}{\alpha_{\rm seg}}
\newcommand{\PLbar}{\overline{\rm PL}}
\newcommand{\sigell}{\sigma_\ell}

\begin{document}

\title{Orbital statistics of multiple systems formed from small-$N$ subclusters
}
\author[H. E. Ambrose, A. P. Whitworth]
{Hannah E. Ambrose$^1$\thanks{E-mail: ambrosehe@cardiff.ac.uk}, A. P. Whitworth$^1$\\
$^1$CHART, School of Physics and Astronomy, Cardiff University, Cardiff CF24 3AA, UK}
\date{Accepted 2025 July 24. Received 2025 April 7; in original form 2025 April 7}

\pubyear{2024}
\label{firstpage}
\pagerange{\pageref{firstpage}--\pageref{lastpage}}
\maketitle

\begin{abstract}
We use numerical $N$-body experiments to explore the statistics of multiple systems formed in small-$N$ subclusters, {\it i.e.} the distributions of orbital semi-major axis, $a$, orbital eccentricity, $e$, mass ratio, $q$, mutual orbital inclination, $\theta$, and ejection velocity, $\upsilon_{\rm ej}$. The stars in a subcluster are evolved as if they are the fragmentation products of a single isolated prestellar core from which most of the natal gas has already been dispersed, and there are no correlations between the stars' initial positions and velocities. Two parameters are particularly important: the number of stars in the subcluster, $N$, and the fraction of kinetic energy in ordered rotation, $\alfROT$. Increasing $N$ has the effect of systematically decreasing the semi-major axes of the tighter orbits, but has very little effect on the semi-major axes of the wider orbits. The main effect of $\alfROT$ is to regulate the distribution of mutual orbital inclinations, with $\alfROT\!\sim\!0.5$ producing a distribution of orbital inclinations for triple systems which is consistent with observed values. Triples frequently form in high-inclination orbits without the assistance of von Zeipel-Lidov-Kozai cycles. Our previous work demonstrated that subclusters with mass segregation, moderate rotation, and typically $N=4$ or 5 stars produced the best fit to the multiplicity statistics (proportions of singles, binaries, triples, etc.). Here we show that these parameters also reproduce the orbital statistics (distributions of orbital semi-major axis, $a$, orbital eccentricity, $e$, mass ratio, $q$, mutual orbital inclination, $\theta$, and ejection velocity, $\upsilon_{\rm ej}$). For the best-fit parameters, $21(\pm 1)\%$ of subclusters produce more than one multiple system.
\end{abstract} 

\begin{keywords}
celestial mechanics -- {\it stars:} binaries {\it (including multiples)}: close -- stars: formation --  stars: kinematics and dynamics
\end{keywords}

\section{Introduction}\label{intro}

Stellar multiples are close, bound systems of stars circling one another on regular, stable or meta-stable orbits. Observational and statistical studies show that such systems are common in the solar neighbourhood. The majority of solar-mass stars appear to be members of multiple systems \citep{2015MNRAS.448.1761W}, and 46\% of solar-mass primaries reside in multiple systems \citep[][hereafter T21]{2021Univ....7..352T}. This percentage increases rapidly with increasing primary mass, reaching $>\!90\%$ for O stars (see \citealt{2023ASPC..534..275O} and references therein). Conversely the percentage decreases with decreasing primary mass. The companion fraction (CF; the mean number of companions per primary star) is found to be highest in the earliest phases of protostellar evolution, and then declines through the subsequent protostellar phases. \cite{2013ApJ...768..110C} observed a CF of $0.91(\pm0.05)$ for low-mass Class 0 protostars, while \cite{2008AJ....135.2496C,2008AJ....135.2526C} found this fraction to be $0.46(\pm0.3)$ for Class 1 sources. \cite{2022ApJ...925...39T} evaluated the same statistics in a more recent, higher resolution study of protostars in Perseus, finding that the CF decreases from $0.74(\pm0.08)$ for Class 0 sources to $0.35(\pm0.09)$ for Class 1. \cite{1993A&A...278...81R} found that the CF continues to fall from pre-Main Sequence to Main Sequence populations.

In \citealt{2024MNRAS.535.3700A} (hereafter AW24), we report the results from a suite of numerical experiments designed to determine the relative numbers of singles, binaries, triples and higher-order multiples (hereafter the {\it Multiplicity Statistics}) formed by pure stellar dynamics in an isolated subcluster with between 3 and 7 members. The experiments that best reproduced the observed Main Sequence multiplicity statistics invoked subclusters with a mix of $N$ values centred on $\mu_N\!\sim\!4.8$, with standard deviation $\sigma_N\!\sim\!2.4$, and moderate ordered rotation, $\alfROT\!\sim\!0.5$. \cite{2013MNRAS.432.3534H} and \cite{LomaxOetal2015} arrived at similar results through independent methods, both concluding that a low-mass prestellar core should typically produce between 4 and 5 stars. \cite{2013MNRAS.432.3534H} reached this conclusion through statistical analysis of the relation between the stellar Initial Mass Function and the Core Mass Function, finding that, if the mapping between the two is self-similar, then $\mu_N\!\sim\!4.3 $, with $\sigma_N\!\sim\!0.4$. \cite{LomaxOetal2015} reached this conclusion by simulating the evolution of Ophiuchus-like prestellar cores using Smoothed Particle Hydrodynamics, finding that $\mu_N\!\sim\!4.5$ and $\sigma_N\!\sim\!1.9$. Recent SPH simulations of molecular clouds by \cite{2025MNRAS.536.3518C} corroborate the \cite{LomaxOetal2015} findings, showing that given a solar neighbourhood star formation rate, as many as 6 stellar-mass fragmentation products will form on the scale of a prestellar core.

Stellar dynamics leaves distinct and lasting signatures on the architectures of multiple systems formed in isolated stellar subclusters. In this regard, an especially important role is played by three-star interactions that start with two of the stars on a relatively tight orbit, and end with two of the stars (not necessarily the same two) on an even tighter orbit. Here we define tighter to mean lower total energy (kinetic plus gravitational), so the third star gains energy (and may be ejected from the subcluster). In such interactions it tends to be the two more massive stars that end up in the tight binary, and this is called {\it Dynamical Biasing} \citep{1993MNRAS.262..800M}.

\cite{1998A&A...339...95S} (hereafter SD98) observe {\it Dynamical Biasing} in their numerical simulations of subclusters with $3\!\leq\!N\!\leq\!5$. Between $80\%$ and $90\%$ of central orbits in triple and higher-order multiples involve the two most-massive stars, and this percentage increases with increasing $N$, and with the range of stellar masses within the subcluster. At the same time, the mean semi-major axis of these central orbits decreases with increasing $N$, since with larger $N$ there are more opportunities for three-star interactions.

In higher-order multiple systems, the orbital inclinations and eccentricities can be altered by von Zeipel-Lidov-Kozai cycles \citep{1910AN....183..345V,1962P&SS....9..719L,1962AJ.....67..591K}. These cycles are caused by a transfer of orbital angular momentum within an hierarchical triple system at each peri-centre of the outer orbit. This transfer results in a cyclic variation in the eccentricity of the inner orbit, combined with a cyclic variation in the relative inclination of the outer orbit.

The dynamics of subclusters also affects the statistics of single stars, since many, and possibly most, single stars start life in subclusters and are then ejected into the field. Consequently the number, mass distribution and velocity distribution of single stars are shaped, at least in part, by the dissolution of their birth subclusters. For example, the single star population should start with a velocity distribution compatible with the distribution of ejection velocities from subclusters. Subsequent interactions between these single stars and other objects (other field stars, star clusters and molecular clouds) may then alter this distribution. In addition, {\it Dynamical Biasing} leads to the preferential ejection of lower-mass stars, which will shift the mass distribution of single stars to lower values than the masses of stars in multiple systems.

In this paper, we analyse the orbital parameters of the multiple systems identified in the numerical experiments of AW24, {\it viz.} semi-major axes; the extent of {\it Dynamical Biasing}; mass ratios; numbers of companions; mutual orbital inclinations; eccentricities; masses and velocities. We term these {\it Orbital Statistics}, to distinguish them from the {\it Multiplicity Statistics} reported in AW24. Section \ref{Method} outlines the model used to initialise and evolve subclusters, and to calculate and monitor their orbital parameters. In Sections \ref{results} and \ref{discussion} we analyse and discuss the results. In Section \ref{conclusion} we summarise our conclusions.

\section{Method} \label{Method}

\subsection{Creating and Evolving Subclusters}\label{SEC:ICs}

In AW24, we evolve an ensemble of stellar subclusters of point-mass particles using pure stellar dynamics, i.e. no ambient gas, therefore no stellar accretion, no stellar mass loss, and no external forces. The subclusters are initially characterised by the number of stars they contain, $N$, and four {\it Configuration Parameters}, namely:
\begin{itemize}
\item $\sigell$, the standard deviation of the log-normal distribution of stellar masses, hereafter termed the `mass range'. Here, $\ell\!=\!\log_{_{10}}\!(M/{\rm M}_{_\odot})$.
\item $\alfROT$, the fraction of kinetic energy in ordered rotation. The rest is in random velocities drawn from an isotropic Maxwellian distribution.
\item $\alfLAW$, the rotation law. This is either Keplerian, $\upsilon_{\rm rot}\propto w^{-1/2}$, or solid-body, $\upsilon_{\rm rot}\propto w$, with $w\!=\!(x^2+y^2)^{1/2}$.
\item $\alfSEG$, which determines whether the stars are mass-segregated ($\alfSEG\!=\!1$) or not ($\alfSEG\!=\!0$).
\end{itemize}
Star positions are generated randomly with uniform density either within a sphere of radius $R_{\rm o}$, or within an oblate spheroid with semi-major axis $R_{\rm o}$.
The procedures used to generate initial particle positions and velocities are detailed in AW24 (SubSection 2.1).\footnote{ In AW24, Section 2.1, we omitted to state how initial star positions were adjusted to flatten rotating configurations. We did not treat this exactly, but simply multiplied the z coordinates of stars by a factor  $F = (1 - \alfROT)/(1 + 2 \alfROT)$, and their $z$ velocity components by  $F^{1/2}$. We note (a) that velocities are subsequently re-normalised to ensure global virial equilibrium, and (b) that we repeated these experiments without flattening and found no significant differences in the results.}

All subclusters start in Virial Equilibrium, and for each combination of $N$ and {\it Configuration Parameters} ($\sigell,\,\alfROT,\,\alfLAW,\,\alfSEG$) we evolve 1000 realisations. The integrator follows a 4th Order Runge-Kutte scheme with global adaptive timestep and no gravitational softening. The median fractional change in the total energy by the end of a simulation is $\sim 1\%$ for $N=3$.
 
 For the purpose of illustration, we set: (i) the mean of the log-normal distribution of masses to $\mu_\ell\!=\!-0.6$ so that the median stellar mass is $M_{\rm med}\!=\!0.25\,{\rm M}_{_\odot}$; and (ii) the radius (or semi-major axis) of the initial spherical (or oblate spheroidal) envelope to $R_{\rm o}\!=\!1000\,{\rm AU}$. However the equations regulating the evolution of a pure $N$-body subcluster are dimensionless, so the results can be rescaled arbitrarily. Specifically, if the total mass of a subcluster is $M_{\rm tot}$, the results can be rescaled to different values of $M_{\rm tot}$ and $R_{\rm o}$ -- say $M_{\rm tot}'$ and $R_{\rm o}'$ -- by multiplying all stellar and system masses by $f_M\!=\!M_{\rm tot}'/M_{\rm tot}$; all position vectors and orbital axes by $f_R\!=\!R_{\rm o}'/R_{\rm o}$; the time and all orbital periods by $(f_R^3/f_M)^{1/2}$; and all velocities by $(f_M/f_R)^{1/2}$. Orbital eccentricities and inclinations are unchanged. Further details can be found in AW24.

\subsection{Identifying and Classifying Multiple Systems}\label{SEC:MMO}

We apply a Multiplicity Monitoring Operation (MMO) at regular intervals (every 33 crossing times) throughout the evolution of a subcluster. 33 crossing times corresponds to $\sim2.3\,{\rm Myr}\,N^{-1/2}$, or $\sim1\,{\rm Myr}$ for a subcluster with $N=4\;{\rm or}\;5$. The MMO determines the full orbital architecture for all multiple systems in the subcluster. At each monitoring timestep, we compute the semi-major axis (a) and eccentricity (e) for all identified orbits, the mutual orbital inclinations ($\theta$) for triples and higher-order multiples, and the masses of the stars and subsystems on either end of the orbit. Further details of the MMO can be found in AW24.

\subsection{Architectures of Multiple Systems}

A system with multiplicity $m$ comprises $m$ stars and $m\!-\!1$ orbits. Most of the parameters we discuss in the sequel describe an orbit. If both the objects on either end of the orbit are stars, we classify the orbit as S2; if only one is a star, as S1; and if neither is a star, as S0.

A binary consists of a single S2 orbit.

A stable (and therefore hierarchical) triple comprises an S2 orbit and a larger S1 orbit; the S2 orbit involves a pair of stars orbiting one another, and the S1 orbit involves this pair and a third star orbiting one another.

A quadruple may comprise an S2 orbit and two S1 orbits, in which case it is classed as a {\it Planetary Quadruple}; the S2 orbit involves a pair of stars orbiting one another, the first S1 orbit involves this pair and a third star orbiting one another, and the second S1 orbit involves this threesome and a fourth star orbiting one another. Alternatively a quadruple may comprise two S2 orbits and one S0 orbit, in which case it is classed as a {\it 2+2 Quadruple}; each of the S2 orbits involves a different pair of stars orbiting one another, and the S0 orbit involves the two pairs orbiting one another.

For the purpose of evaluating statistical distributions we distinguish S2 orbits from S1 orbits, and we do not consider S0 orbirts, since they are too rare to support reliable statistics.

\section{Results}\label{results}

We have selected two main cases from AW24 with which to illustrate the {\it Orbital Statistics}. In both we fix $\sigma_\ell\!=\!0.3$, since this is the value that is most consistent with the mapping from the prestellar Core Mass Function to the stellar Initial Mass Function (Whitworth, Ambrose \& Georgatos, in prep.).

\begin{itemize}

\item The first case is the {\it Fiducial Case}. It has $\alfROT\!=\!0$, therefore there is no ordered rotation and $\alfLAW$ is irrelevant. It also has $\alfSEG\!=\!0$, meaning no mass segregation. We present results for $N\!=\!3$, $4$, $5$, $6$ and $7$. With $N\!=\!1$ all stars are single, and with $N\!=\!2$ all stars are in binaries, since the subcluster is virialised from the outset.

\item The second case is the \textit {Best-Fit Case}. This is the case identified in AW24 which best reproduces the {\it Multiplicity Statistics} for solar-mass primaries from T21. It has a distribution of $N$ values with mean $\mu_N\!=\!4.8$ and standard deviation $\sigma_N\!=\!2.4$. 
The Configuration Parameters are $\alfROT=0.5$ and $\alfLAW\!=\!\mbox{\sc kep}$ (moderate Keplerian rotation) and $\alfSEG\!=\!1$ (mass segregation).
The procedure used to identify the \textit {Best-Fit Case} is detailed in AW24 Section 3.2 and SubSection 3.2.1.

\item Dependences that are better illustrated by considering parameter values other than those of the {\it Fiducial} and {\it Best-Fit Cases} are discussed in separate subsections. Section \ref{SEC:HighRot.01} deals with the dependence of semi-major axes on the amount of ordered rotation  ($\alfROT$). Section \ref{sec:arot} deals with how the mutual orbital inclinations depend on the amount of ordered rotation  ($\alfROT$).

\end{itemize}

\subsection{The Distribution of Semi-Major Axes}

\begin{figure*}
\centering
\includegraphics[width=0.75\textwidth]{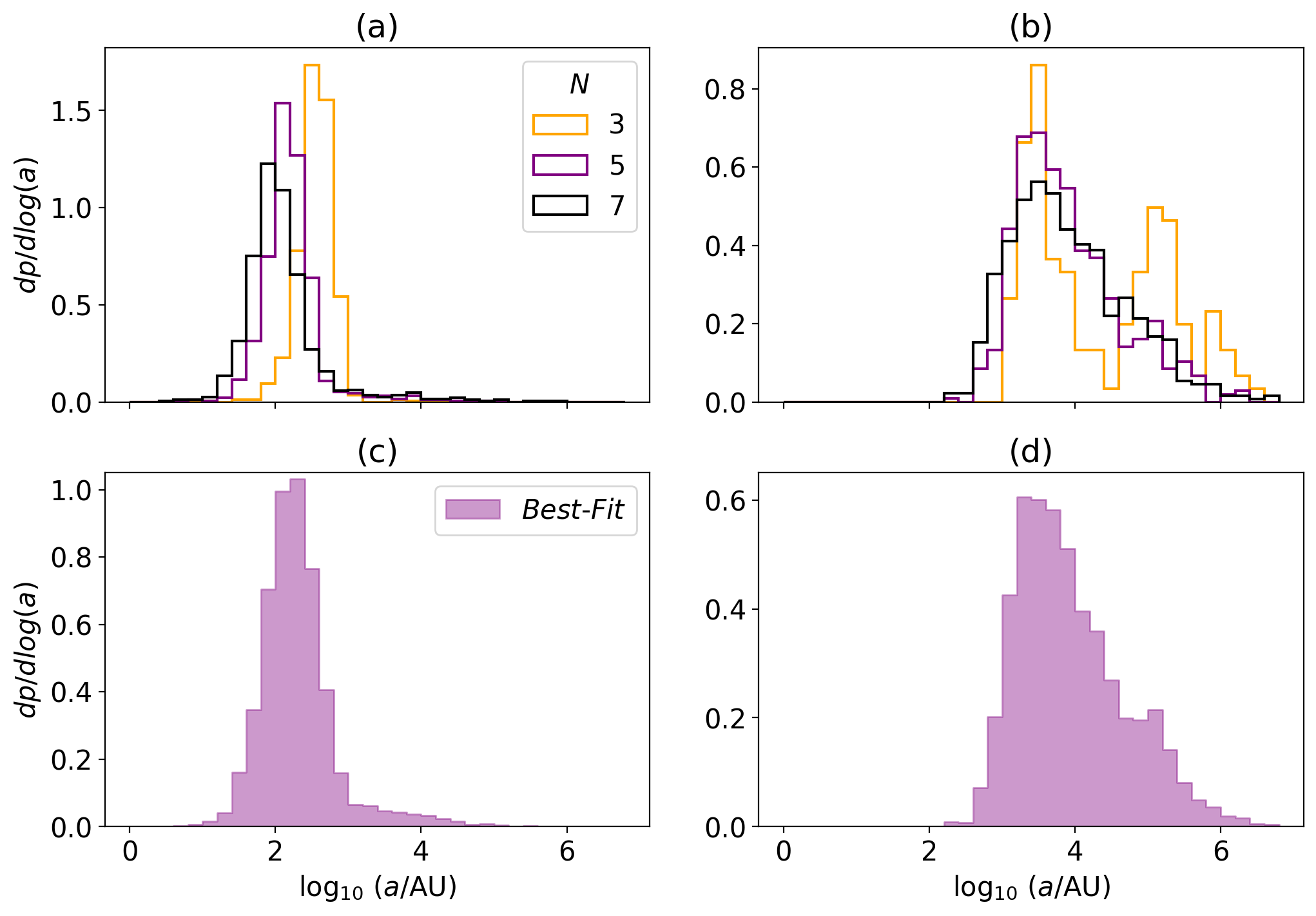}
\hfill    
\caption{The specific probability distributions of semi-major axis, $a$, for (a) S2 orbits, and (b) S1 orbits, in the {\it Fiducial Case} with $N\!=\!3$, $5$ and $7$; the results for $N\!=\!4$ and $6$ are omitted to avoid confusion. (c) and (d) show the specific probability distributions for the {\it Best-Fit Case}.} 
\label{fig:aComp}
\end{figure*}

\begin{figure}
\centering
\includegraphics[width=0.48\textwidth]{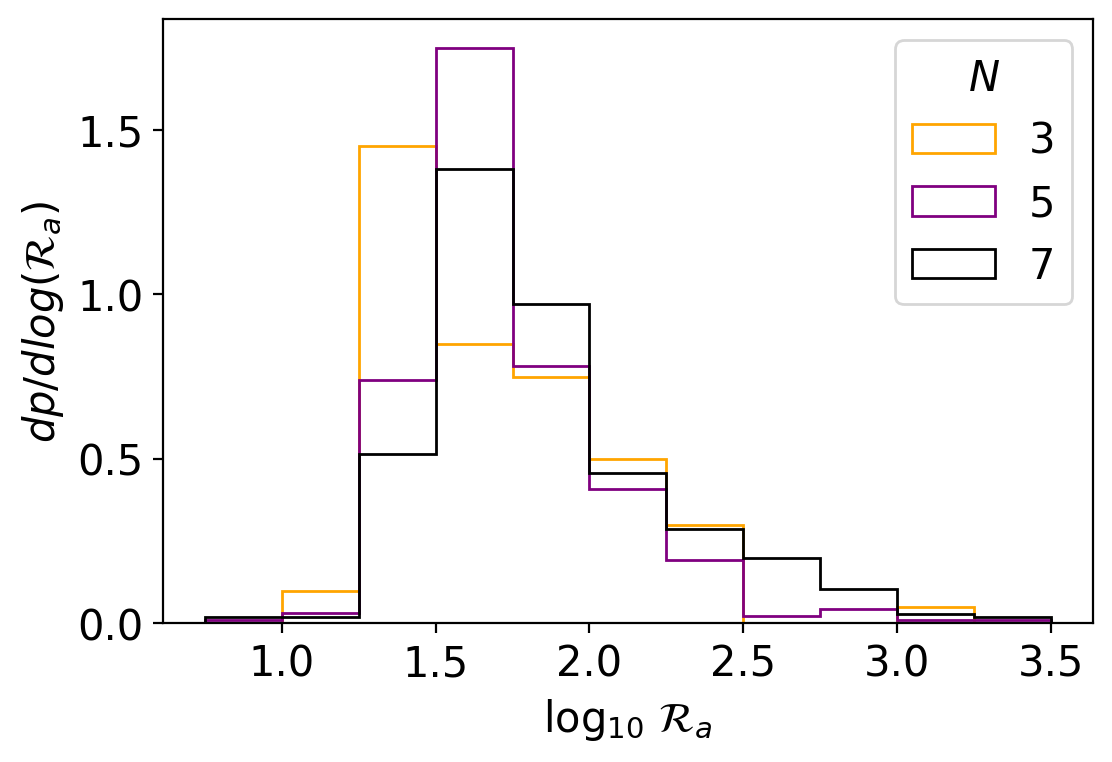}
\hfill    
\caption{The specific probability distributions of the ratio, $\mathcal{R}_a$, of outer (S1) to inner (S2) orbital semi-major axis in triple systems for $N=3$, 5, and 7.}
\label{fig:aratio}
\end{figure}

\begin{figure}
\centering
\includegraphics[width=0.48\textwidth]{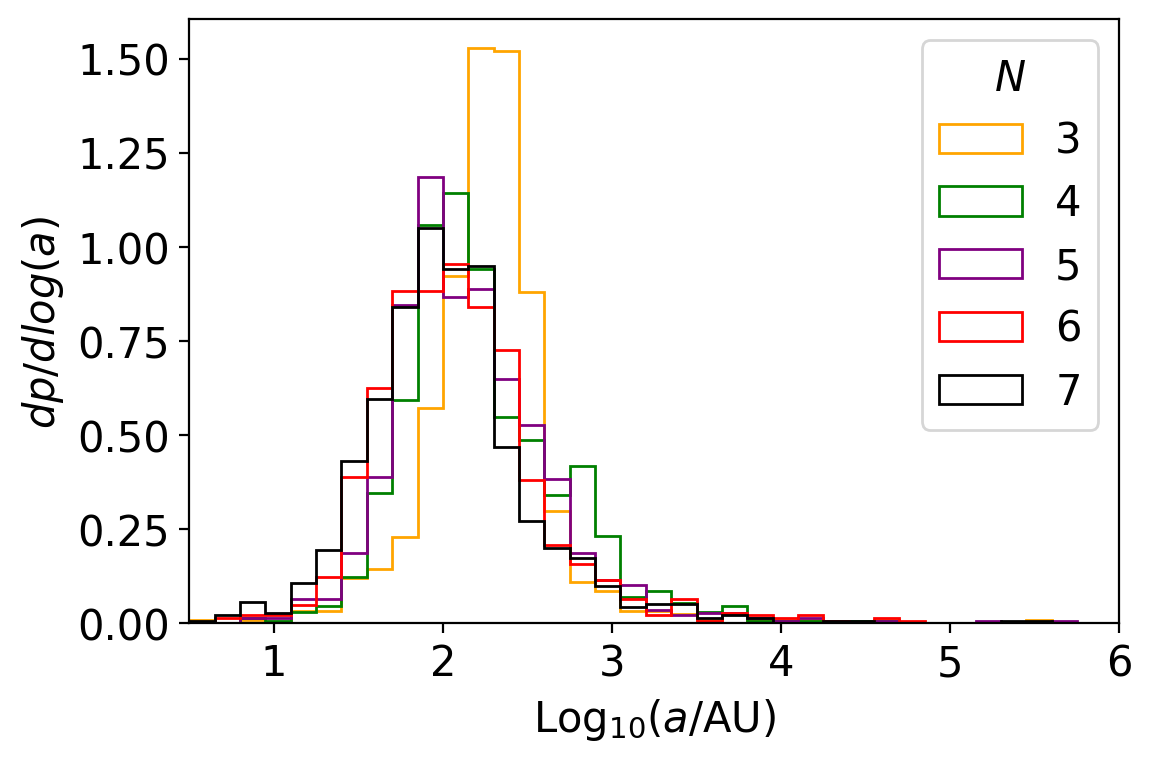}
\hfill    
\caption{The specific probability distributions of semi-major axis, $a$, for S2 orbits from the high-rotation case discussed in Section \ref{SEC:HighRot.01} (i.e. $\alfROT\!=\!0.99$, $\alfLAW\!=\!\mbox{\sc sol}$, $\alfSEG\!=\!0$, so almost all the kinetic energy in solid-body rotation, and no mass segregation) ~with $N\!=\!3$, $4$, $5$, $6$, $7$.}
\label{fig:aB99}
\end{figure}

\subsubsection{Semi-major axes in the Fiducial Case}\label{SEC:SMAfiducial}

Figures \ref{fig:aComp}a and \ref{fig:aComp}b show that in the {\it Fiducial Case} the semi-major axes of S2 orbits tend to decrease with $N$, whereas the semi-major axes of S1 orbits are roughly independent of $N$. Specifically, with the scalings adopted here (see Section \ref{SEC:ICs}), the peak of the distribution of semi-major axes for S2 orbits decreases from $\sim\!400\,{\rm AU}$ for $N\!=\!3$, to $\sim\!90\,{\rm AU}$ for $N\!=\!7$, with the peaks of the corresponding period distributions decreasing from $\sim\!8\,{\rm kyr}$ to $\sim\!1\,{\rm kyr}$. In contrast, the semi-major axes for S1 orbits peak at $\sim\!4,000\,{\rm AU}$ (periods at $\sim\!300\,{\rm kyr}$), more or less independent of $N$. The ratios of S1 to S2 orbital semi-major axes, $\mathcal{R}_a$, are shown in Figure \ref{fig:aratio}. The ratio peaks around $\mathcal{R}_a \sim45$ for $N=7$ and drops to $\mathcal{R}_a \sim25$ for $N=3$.

\subsubsection{Semi-major axes in the Best-Fit Case}\label{SEC:SMAbestfit}

Figures \ref{fig:aComp}c and \ref{fig:aComp}d show that in the {\it Best-Fit Case}, the distribution of semi-major axes for S2 orbits peaks at $\sim\!160\,{\rm AU}$ (periods $\sim\!2.5\,{\rm kyr}$), similar to the {\it Fiducial Case} with $N\!=\!4$ or $5$. For S1 orbits the distribution peaks at $\sim\!4,000\,{\rm AU}$ (periods $\sim\!300{\rm kyr}$), as for the {\it Fiducial Case} with all $N$.

\subsubsection{Semi-Major axes with Very High Rotation} \label{SEC:HighRot.01}

The distributions of semi-major axis for S2 orbits (as discussed in the two preceding subsections, \ref{SEC:SMAfiducial} and \ref{SEC:SMAbestfit}) are, with one notable exception, almost independent of the Configuration Parameters ($\sigma_\ell$, $\alfROT$. $\alfLAW$, $\alfSEG$). The one exception is the case of very rapid rotation. Figure \ref{fig:aB99} shows the results obtained with ~$\sigma_\ell\!=\!0.3$, ~$\alfROT\!=\!0.99$, $\alfLAW\!=\!\mbox{\sc sol}$, ~$\alfSEG\!=\!0$, ~and different $N$. These parameters correspond to the same range of un-segregated masses as \ref{SEC:SMAfiducial} and \ref{SEC:SMAbestfit}, but with almost all the kinetic energy invested in solid-body rotation. In this case the semi-major axes again tend to decrease with increasing $N$, but the effect is significantly smaller. There are two reasons for this. First the stars start on very circular orbits, and therefore close to nested `planetary' architectures; this reduces the frequency of close interactions. Second, with solid-body rotation, and a gravitational potential well that is only due to the masses of the other stars, the Virial Condition in the form
\begin{eqnarray}
E_{\rm kinetic}&=&E_{\rm random}\,+\,E_{\rm rotation}\;\;\;=\;\;\;-\,\frac{\Omega}{2}\,,\hspace{0.7cm}
\end{eqnarray} 
will often deliver setups in which some stars are unbound, from the outset, and therefore fly off immediately. At the same time, the stars on closer orbits are very strongly bound and tend to stay so. This is particularly true for $N\!=\!3$.

\subsection{Mass Ratios} \label{sec:MR}

\begin{figure}
\centering
\includegraphics[width=0.48\textwidth]{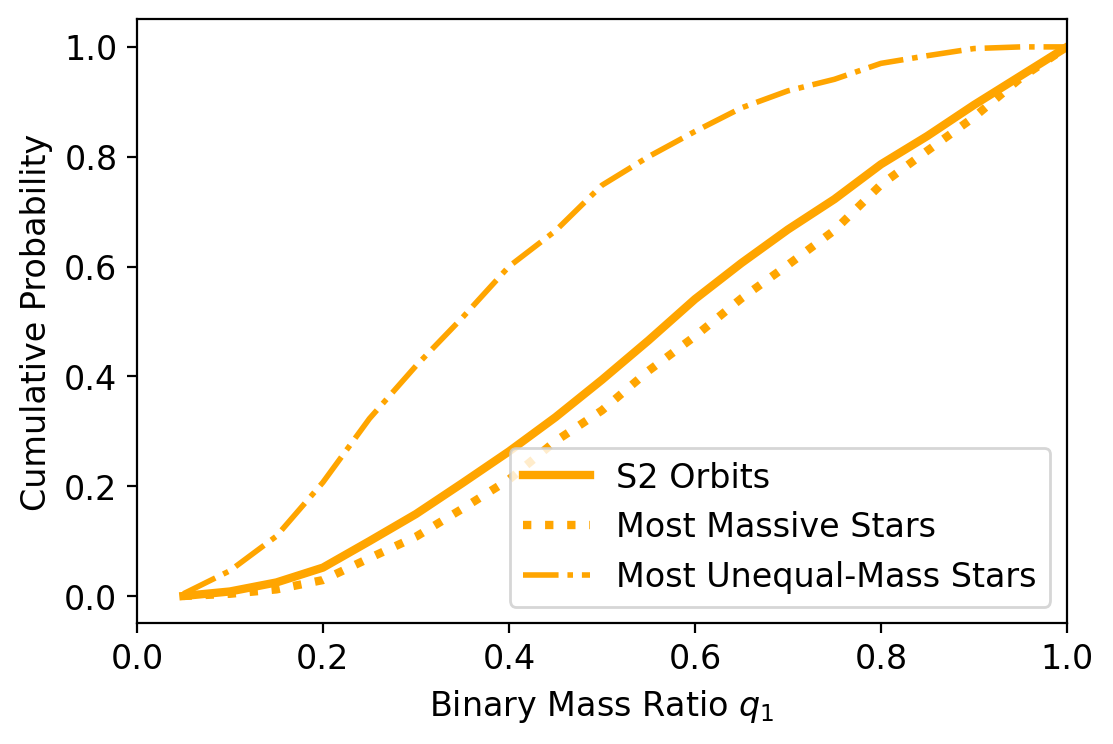}
\hfill    
\caption{The cumulative probability distributions of mass ratio, $q_{_1}$. ~{\it Solid Line}: S2 orbits of systems formed dynamically in the {\it Fiducial Case} with $N = 3$. -{\it Dotted Line:} values obtained by simply pairing the 2 most massive stars. -{\it Dash-Dotted Line:} values obtained by pairing the most and least massive stars.}
\label{fig:BinMRKS3}
\end{figure}

\begin{figure*}
\centering
\includegraphics[width=0.75\textwidth]{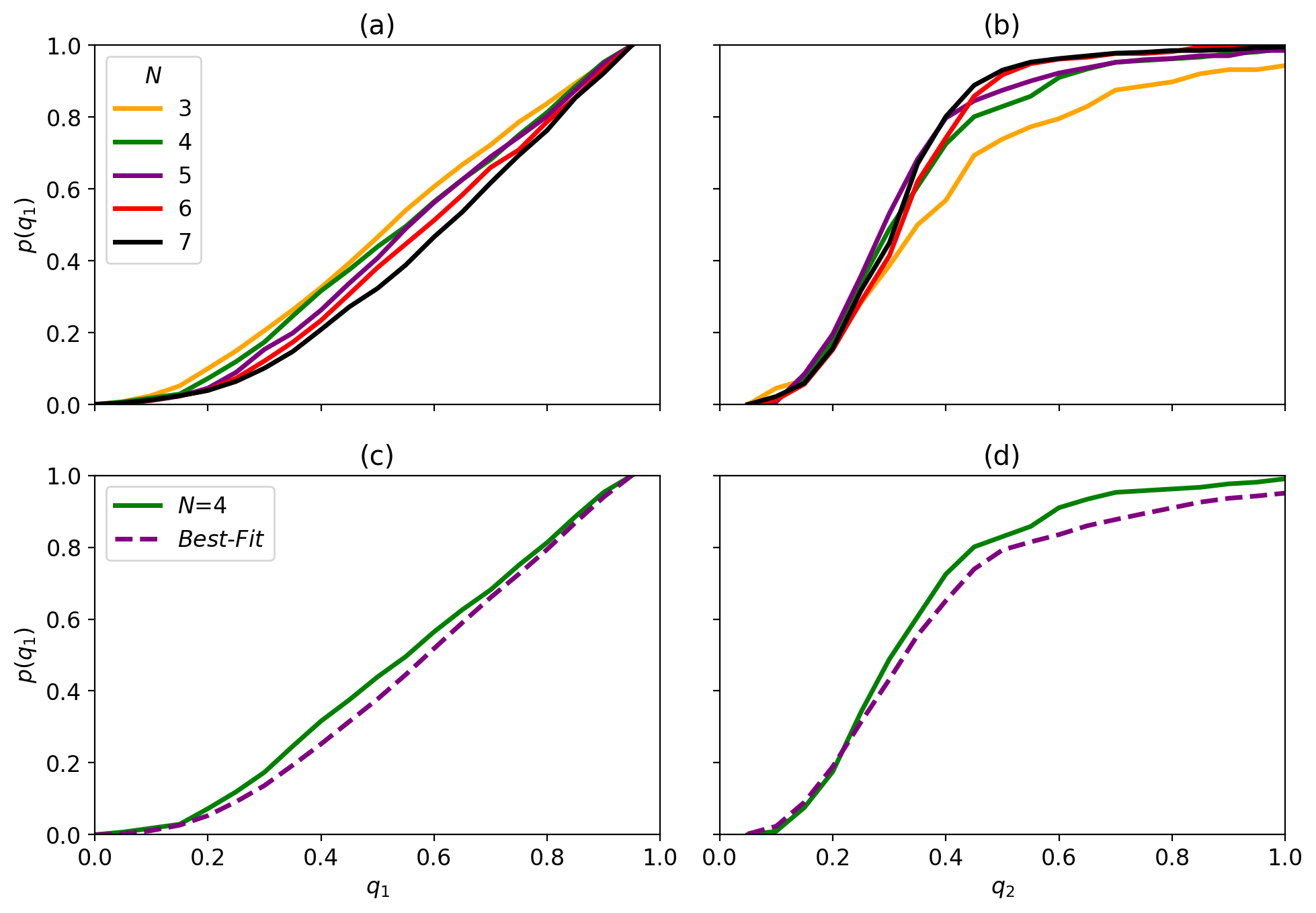}
\hfill    
\caption{The cumulative probability distributions of binary mass ratio, $q_{_1}$, and tertiary mass ratio, $q_{_2}$. ~(a) $q_{_1}$, and (b) $q_{_2}$, for the {\it Fiducial Case} with $N\!=\!3$, $4$, $5$, $6$, $7$. ~(c) $q_{_1}$, and (d) $q_{_2}$, for the {\it Best-Fit Case} with {\it Fiducial Case} $N=4$ for comparison.}
\label{fig:BinMR}
\end{figure*}

\subsubsection{Mass Ratios in the Fiducial Case} \label{sec:Qfid}

For any system in which two stars with masses $M_{_1}$ and $M_{_2}\,(<\!M_{_1}$) orbit their mutual centre of mass, the binary mass ratio is $q_{_1}\!=\!M_{_2}/M_{_1}$. Necessarily $0\!<\!q_{_1}\!<\!1$.

For any system in which a star with mass $M_{_3}$ and an S2 pairing (total mass $M_{_1}\!+\!M_{_2}$) orbit one another, the tertiary mass ratio is $q_{_2}\!=\!M_{_3}/(M_{_1}\!+\!M_{_2})$. It follows that $q_{_2}\!>\!0$, but there is no upper limit on $q_{_2}$.

The solid line on Figure \ref{fig:BinMRKS3} shows the cumulative probability distribution of $q_{_1}$ for the S2 orbits from the {\it Fiducial Case} with $N\!=\!3$. For comparison, the dotted line shows the ratio between the mass of the second most massive star and the mass of the most massive star, irrespective of whether they end up on an S2 orbit, for the same case. The dash-dotted line shows the ratio between the mass of the most and least massive stars. This demonstrates the tendency of pure $N$-body dynamics to deliver the two most massive stars into an S2 orbit (and conversely to put less massive stars on outer orbits or eject them). In this case $72(\pm1.5)\%$ of the S2 orbits involve the two most massive stars.

Figure \ref{fig:BinMRKS3} also shows that the distribution of mass ratios for the S2 orbits is approximately flat in the interval $0.2\!<\!q_{_1}\!<\!1$, and that there are very few below $q_{_1}\!=\!0.2$, i.e. to a first approximation,
\begin{eqnarray}
\frac{dp}{dq_{_1}}&\sim&\left\{\begin{array}{ll}
0.00\,,&q_{_1}\!<\!0.2\,;\\
1.25\,,\hspace{0.4cm}&0.2\!\leq\!q_{_1}\!<\!1.\\
\end{array}\right.
\end{eqnarray}

\begin{table}
\caption{The median values of binary mass ratio, $q_{_{1,\textrm{med}}}$, for all $N$ values in the \textit{Fiducial Case}, and the \textit{Best-Fit Case}.}
\begin{center}
\begin{tabular}{lcccccc}
$N$ & 3 & 4 & 5 & 6 & 7 & \textit{Best-Fit}  \\ \hline
$q_{_{1,\textrm{med}}}$ & 0.58 & 0.62 & 0.63 & 0.65 & 0.66 & 0.60  \\ 
\end{tabular}
\end{center}
\label{tab:q1med}
\end{table}

Figure \ref{fig:BinMR}a shows the cumulative probability distribution of $q_{_1}$ for the S2 orbits in the {\it Fiducial Case} with $N\!=\!3$, $4$, $5$, $6$, $7$. As $N$ increases, the $q_{_1}$ distribution for S2 orbits shifts to higher values.\footnote{While mass ratios and the relative frequency of high-$q_1$ systems increases with $N$, the degree of pure {\it Dynamical Biasing}, in which S2 orbits involve the two most massive stars, actually decreases with increasing $N$. This is because, with larger $N$, i.e. more stars in the birth subcluster and therefore more relatively massive stars to chose from, a pairing of -- say -- the first and third most massive star may still have a high mass ratio. As noted by SD98, {\it Dynamical Biasing} is a more useful concept if it is not limited in this sense, i.e. not limited to systems involving the two most massive stars.} This shift is apparent from the median values of $q_1$, shown in Table \ref{tab:q1med}.

Our results for the {\it Fiducial Case} and $N\!=\!3$, $4$, $5$ can be compared with those of SD98, who obtain somewhat higher levels of dynamical biasing. This is because SD98 initialise their subclusters with a greater spread of stellar masses.

Our model almost never produces binaries with very similar masses, say $q_{_1}\!>\!0.9$, since this requires two unlikely circumstances: the random selection of two stars with very similar mass from the mass spectrum, {\it and} for those two stars to be high-mass and therefore likely to pair up.

Figure \ref{fig:BinMR}b shows the cumulative probability distribution of $q_{_2}$ for the S1 orbits in the {\it Fiducial Case} with $N\!=\!3$, $4$, $5$, $6$, $7$. ~As $N$ increases, the mean $q_{_2}$ decreases, and in the limit $N\!\gtrsim\!5$ almost all higher-order systems have $0.1\,\lesssim\!q_{_2}\!\lesssim\!0.4$. For lower $N$ values there is little ($N\!=\!4$) or no ($N\!=\!3$) choice for the masses of additional components in higher-order systems, once the stars in the central S2 orbit have been set.

\subsubsection{Mass Ratios in the Best-Fit Case}

Figure \ref{fig:BinMR}c shows the cumulative probability distribution of $q_{_1}$, and Figure \ref{fig:BinMR}d the cumulative probability distribution of $q_{_2}$, for the {\it Best-Fit Case}. They are very similar to those for the {\it Fiducial Case} with $N\!=\!4$ and $N\!=\!5$, {\it i.e.} most $q_{_1}$ values are between $0.2$ and $0.9$; ~most $q_{_2}$ values are between $0.1$ and $0.4$; ~$65\%$ of S2 orbits involve the two most massive stars in the initial subcluster. Rotation and mass-segregation do not have a significant influence on mass ratios.

\subsection{The Number of Companions}

\begin{figure}
\centering
\includegraphics[width=0.48\textwidth]{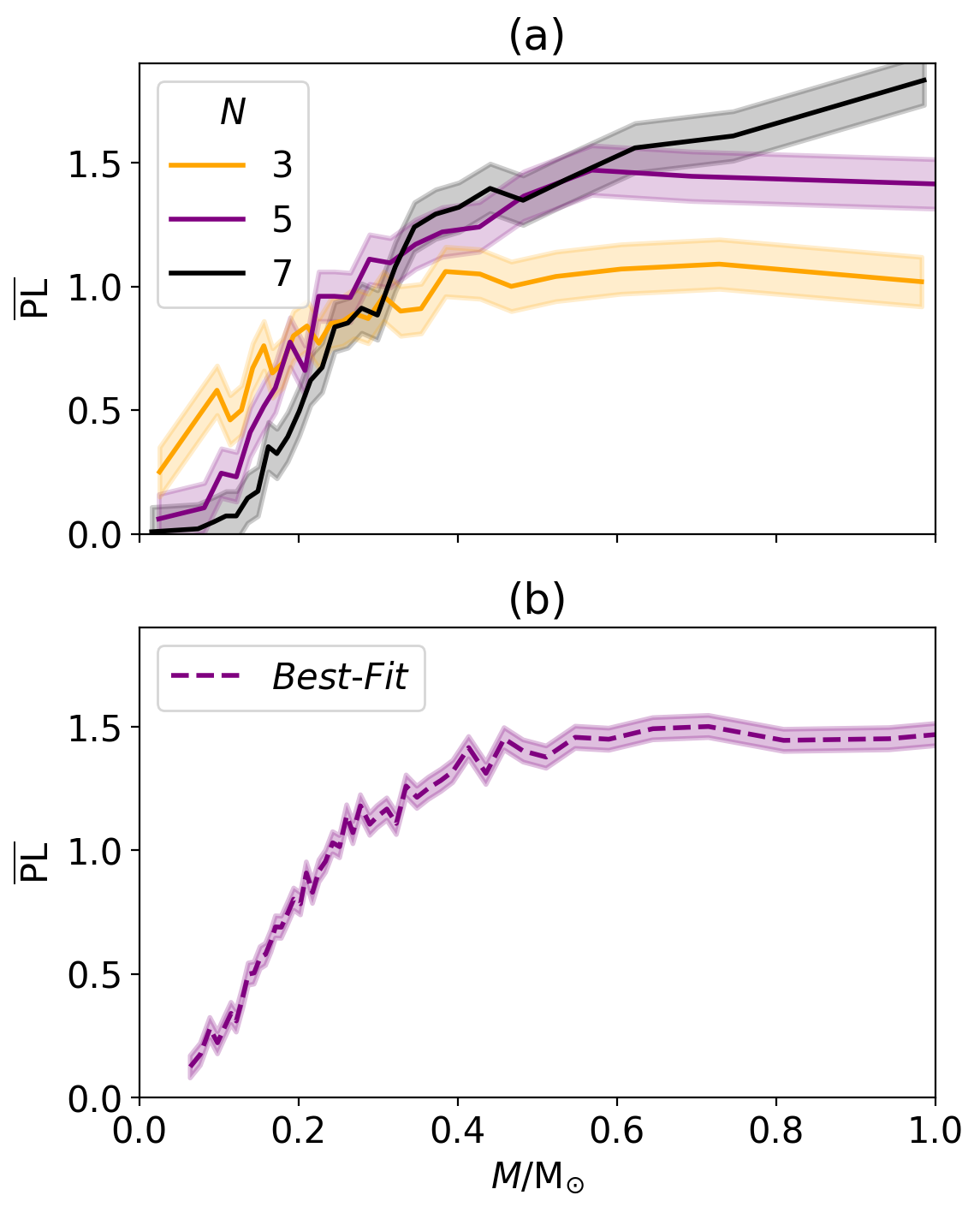}
\hfill    
\caption{The mean plurality, $\overline{\rm PL}$, as a function of stellar mass. ~(a) the {\it Fiducial Case} with $N\!=\!3$, $5$, $7$. ~(b) The {\it Best-Fit Case}. The shaded regions represent the $1\sigma$ uncertainty.}
\label{fig:avecomp}
\end{figure}

In AW24, we define the Plurality of a cohort of stars,
\begin{eqnarray}\label{EQN:PL}
\rm PL&=&\rm\frac{2B+6T+12Q+...}{S+2B+3T+4Q+...}\,,
\end{eqnarray}
which is the mean number of companions that a star in the cohort has, {\it irrespective of whether it is a primary star}. Thus, for example, a binary involves two stars each of which has one companion, a triple involves three stars each of which has two companions, and so on.

\subsubsection{The Number of Companions in the Fiducial Case} \label{sec:avecompfid}

Figure \ref{fig:avecomp}a shows the mean Plurality, $\PLbar$, as a function of mass for the {\it Fiducial Case} with $N\!=\!3$, $5$, $7$. As expected $\PLbar$ increases with mass, and exceeds unity for masses above the median ($0.25\,{\rm M}_{_\odot}$). As $N$ increases, the mean Plurality of the highest-mass stars increases, and the mean Plurality of the lowest-mass stars decreases. This is a consequence of the greater number of interactions that can occur when $N$ is higher. These interactions tend to eject lower-mass stars, thereby reducing their Plurality, and to deliver higher-mass stars into more tightly-bound long-lived higher-order multiples, thereby increasing their Plurality.

\subsubsection{The Number of Companions in the Best-Fit Case}

Figure \ref{fig:avecomp}b shows the variation of $\PLbar$ with stellar mass for the {\it Best-Fit Case}. This is very similar to the {\it Fiducial Case} with $N\!=\!5$, indicating that rotation and mass segregation do not affect $\PLbar$ very much. For masses $M\gtrsim0.5\,{\rm M}_{_\odot}$, $\,\PLbar\simeq1.5$.

\subsection{Mutual Orbital Inclination and von Zeipel-Lidov-Kozai Cycles}

\begin{figure}
\centering
\includegraphics[width=0.48\textwidth]{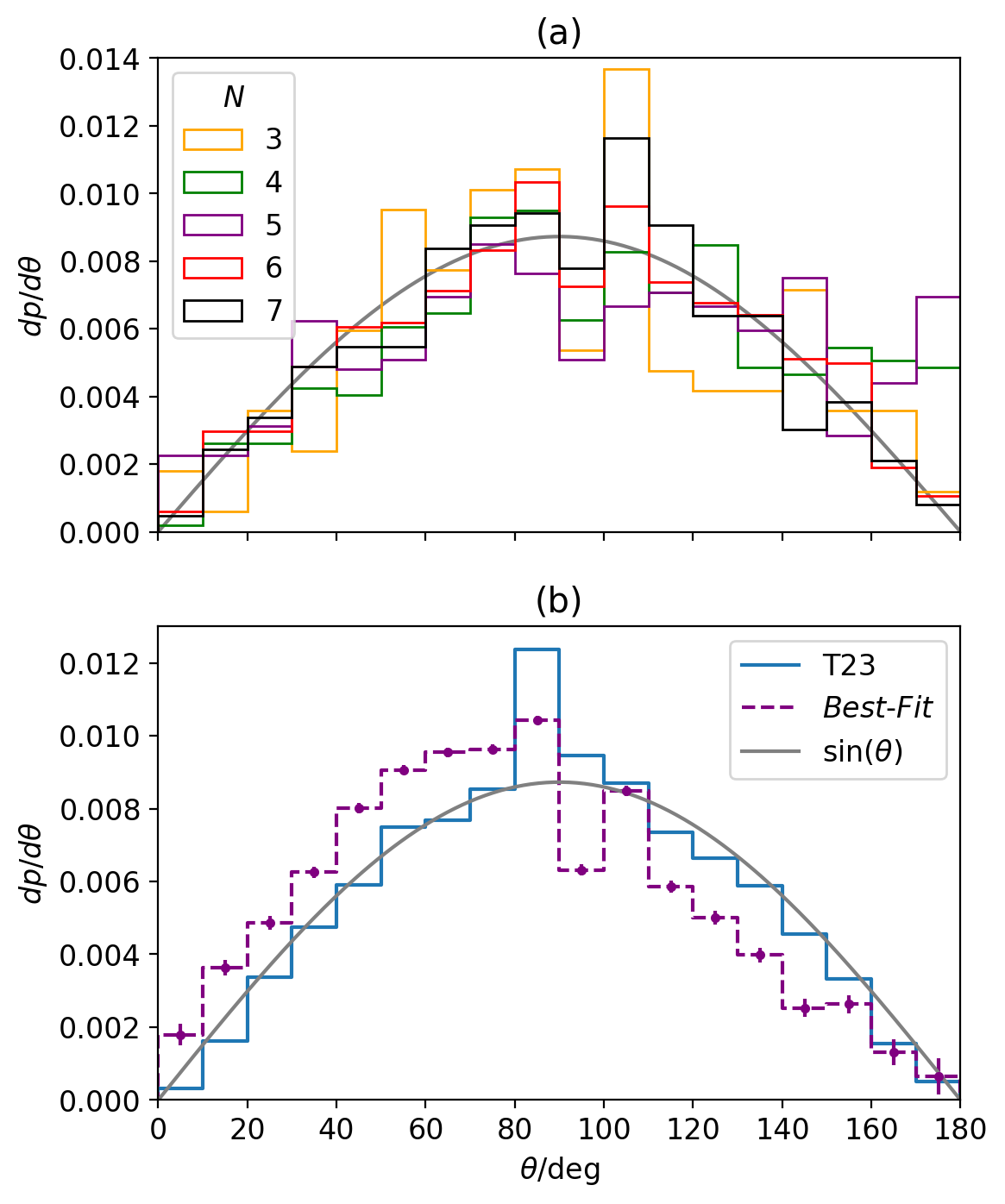}
\hfill    
\caption{The specific probabiity distributions of mutual orbital inclination, $\theta$, for $\theta_o$ in degrees. ~(a) The {\it Fiducial Case} with $N\!=\!3$, $4$, $5$, $6$, $7$. ~(b) The {\it Best-Fit Case} and for the observed systems in the T23 catalogue. Figure \ref{fig:iorbB0}a is produced using 2000 realizations for each $N$ value, in order to improve signal-to-noise.}

\label{fig:iorbB0}
\end{figure}

\begin{figure}
\centering
\includegraphics[width=0.48\textwidth]{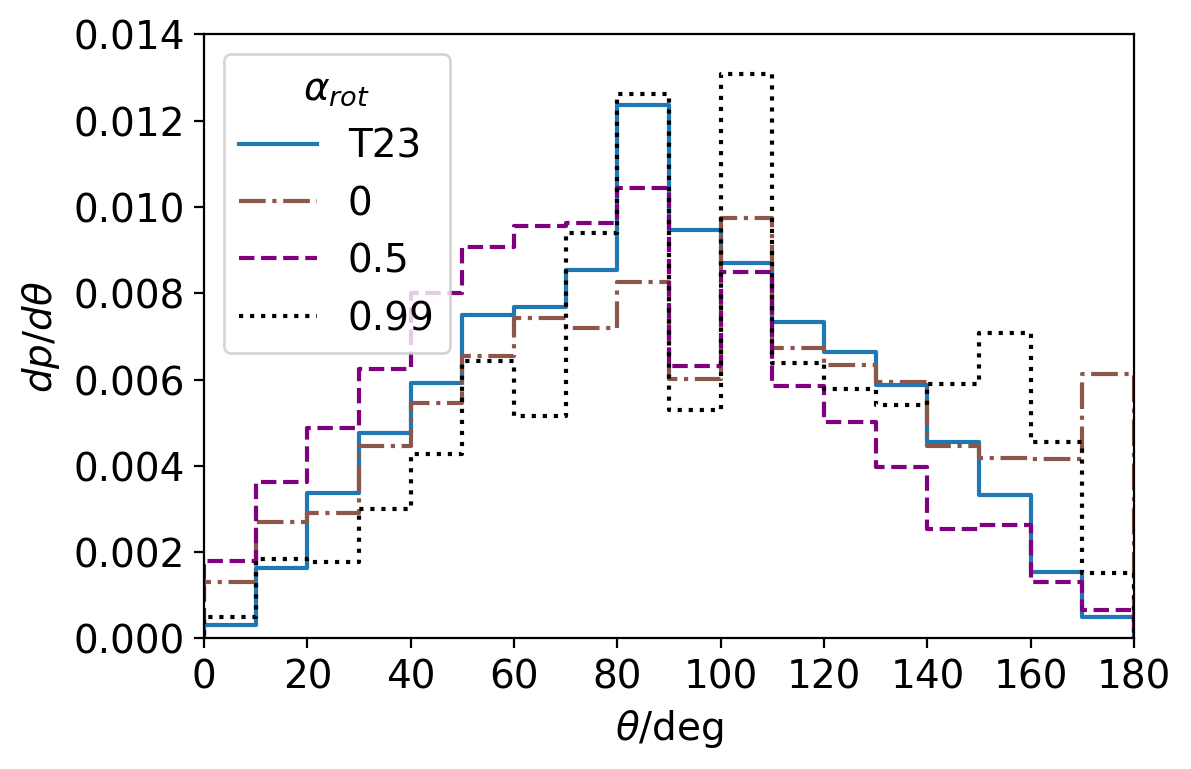}
\caption{The cumulative probability distributions of mutual orbital inclination, $\theta$, for subclusters with the {\it Best-Fit} $N$ and different amounts of initial solid-body rotation, $\alfROT$; and for the observed systems in the T23 catalogue.}
\label{fig:Bcomp}
\end{figure}

\begin{figure}
\centering
\includegraphics[width=0.48\textwidth]{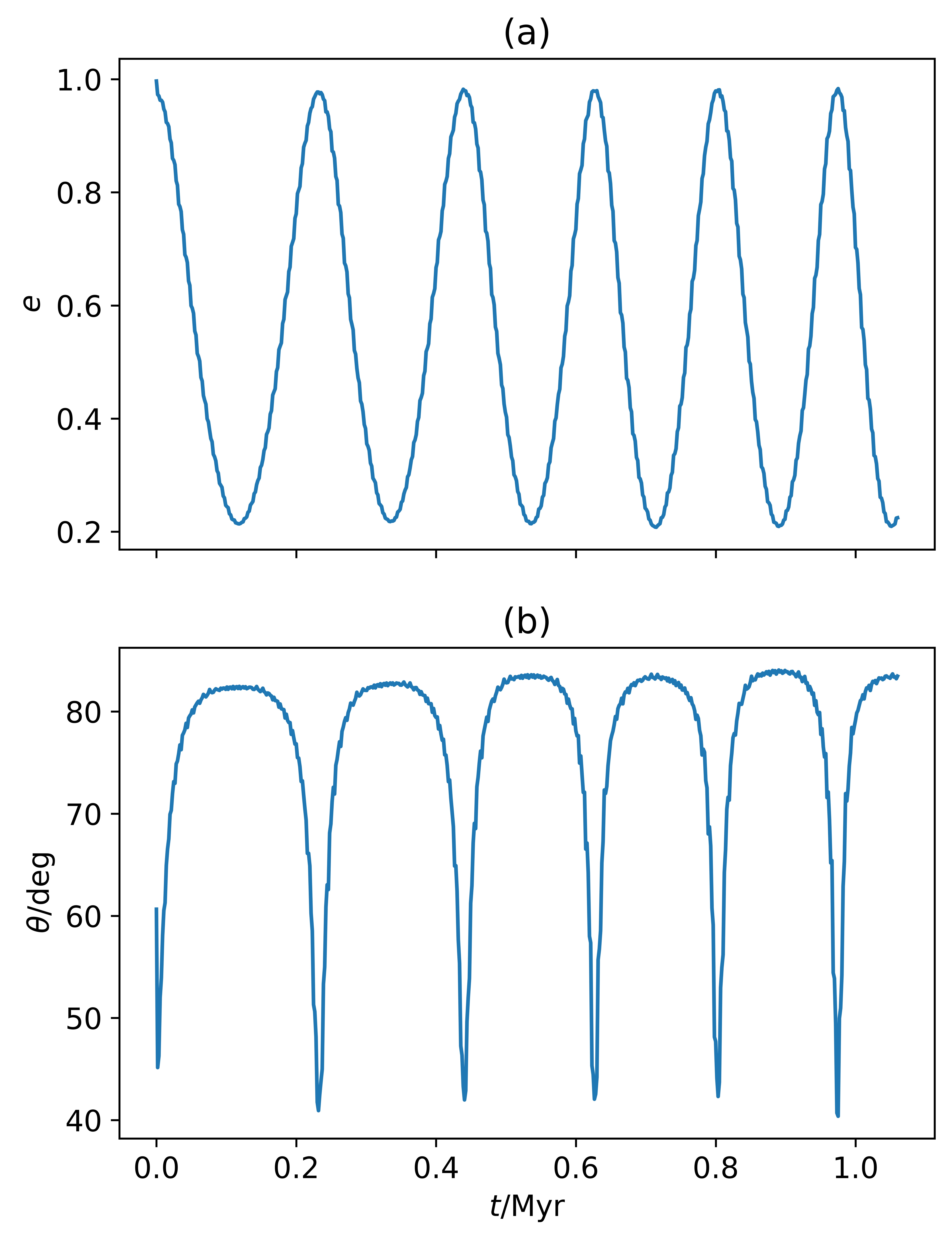}
\hfill  
\caption{An hierarchical triple system undergoing von Zeipel-Lidov-Kozai cycles. (a) The eccentricity, $e$, of the inner S2 orbit, as a function of time. (b) The corresponding mutual orbital inclination, $\theta$.}
\label{fig:kozai}
\end{figure}

Each orbit, $o$, in a multiple system has an orientation, $\hat{\boldsymbol e}_o={\boldsymbol L}_o/|{\boldsymbol L}_o|$, which is the direction of the associated angular momentum, ${\boldsymbol L}_o$. An important parameter constraining the architectures of triple and higher-order multiples is the mutual orbital inclination, i.e. the angle $\theta_{oo'}\!=\!\cos^{-1}(\hat{\boldsymbol e}_o\cdot\hat{\boldsymbol e}_o')$ between two nested orbits $o$ and $o'$. A random, isotropic distribution of orbits will have a mutual orbital inclination PDF with $dp/d\theta\!= \!\sin(\theta)/2$, where $0 \leq \theta \leq \pi$. In this section we explore the statistics of mutual orbital inclinations, $\theta$, and their systematic time-variation, due to von Zeipel-Lidov-Kozai cycles.

\subsubsection{Mutual orbital inclination in the Fiducial Case}

Figure \ref{fig:iorbB0}a shows the specific probability distributions of mutual orbital inclinations for the {\it Fiducial Case} with different $N$. For all $N$ there are very few systems close to co-rotation (small $\theta$). For $N\!=\!3$, $6$, and $7$, the distribution is peaked towards approximately orthogonal orbits. For $N\!=\!4$ and $5$, the distribution is quite flat above $\theta\!=\!30^{\rm o}$, implying that counter-rotating orbits are somewhat favoured compared with a random $\theta$ distribution. This preference for counter-rotation is supported by the simulations of \cite{2022ApJ...939...81H}, who find hierarchical triples to be more stable in an orbit that is fully retrograde than either fully prograde or orthogonal.

\subsubsection{Mutual orbital inclination in the Best-Fit Case}

Figure \ref{fig:iorbB0}b shows the specific probability distributions of mutual orbital inclinations for the {\it Best-Fit Case} and for the observed systems in the \citealt{vizier:J/ApJS/235/6} catalogue (hereafter T23). They have similar shapes, but in the {\it Best-Fit Case} the mean is somewhat lower ($\bar{\theta}\!=\!78^\circ$ with skewness $0.038$) than the T23 sample ($\bar{\theta}\!=\!89^\circ$ with skewness $0.035$). The dip in orbits with $\theta \sim 90^\circ$ is also supported by the \cite{2022ApJ...939...81H} results, which find initially-orthogonal orbits to be the least stable when compared with fully prograde and fully retrograde.

\subsubsection{The Effect of Rotation} \label{sec:arot}

Figure \ref{fig:Bcomp} shows the cumulative probability distributions of mutual orbital inclinations for the systems in the T23 catalogue, and the effect of introducing different amounts of ordered rotation, $\alfROT$, into the {\it Best-Fit Case}. There is a better fit when $\alfROT\!=\!0.5$ (albeit $\bar{\theta}$ is a little low) than with very low rotation ($\alfROT\!=\!0$) or very high rotation ($\alfROT\!=\!0.99$), which both produce distributions that flatten toward high inclinations ($\theta>110^{\circ}$).

\subsubsection{von Zeipel-Lidov-Kozai Cycles}

In a von Zeipel-Lidov-Kozai (ZLK) cycle, a slow periodic change in the eccentricity of the inner binary produces a correlated change in the inclination of the tertiary orbit at each peri-centre. An example from our numerical experiments is shown in Figure \ref{fig:kozai}. In the case of hierarchical triples with high eccentricity outer orbits, octupole-order interactions can drive the Eccentric Kozai-Lidov (EKL) mechanism (see \citealt{2016ARA&A..54..441N}), potentially flipping the orbit from $\theta\!<\!90^\circ$ to $\theta\!>\!90^\circ$. However, of the systems that end up with $\theta\!>\!90^\circ$ in the experiments reported here, only $\sim\!25\%$ reach these inclinations by the EKL mechanism. The remainder are either formed with, or displaced impulsively into, orbits with $\theta\!>\!90^\circ$.

\subsection{Orbital Eccentricities}\label{SEC:OrbEcc}

\begin{figure*}
\centering
\includegraphics[width=0.75\textwidth]{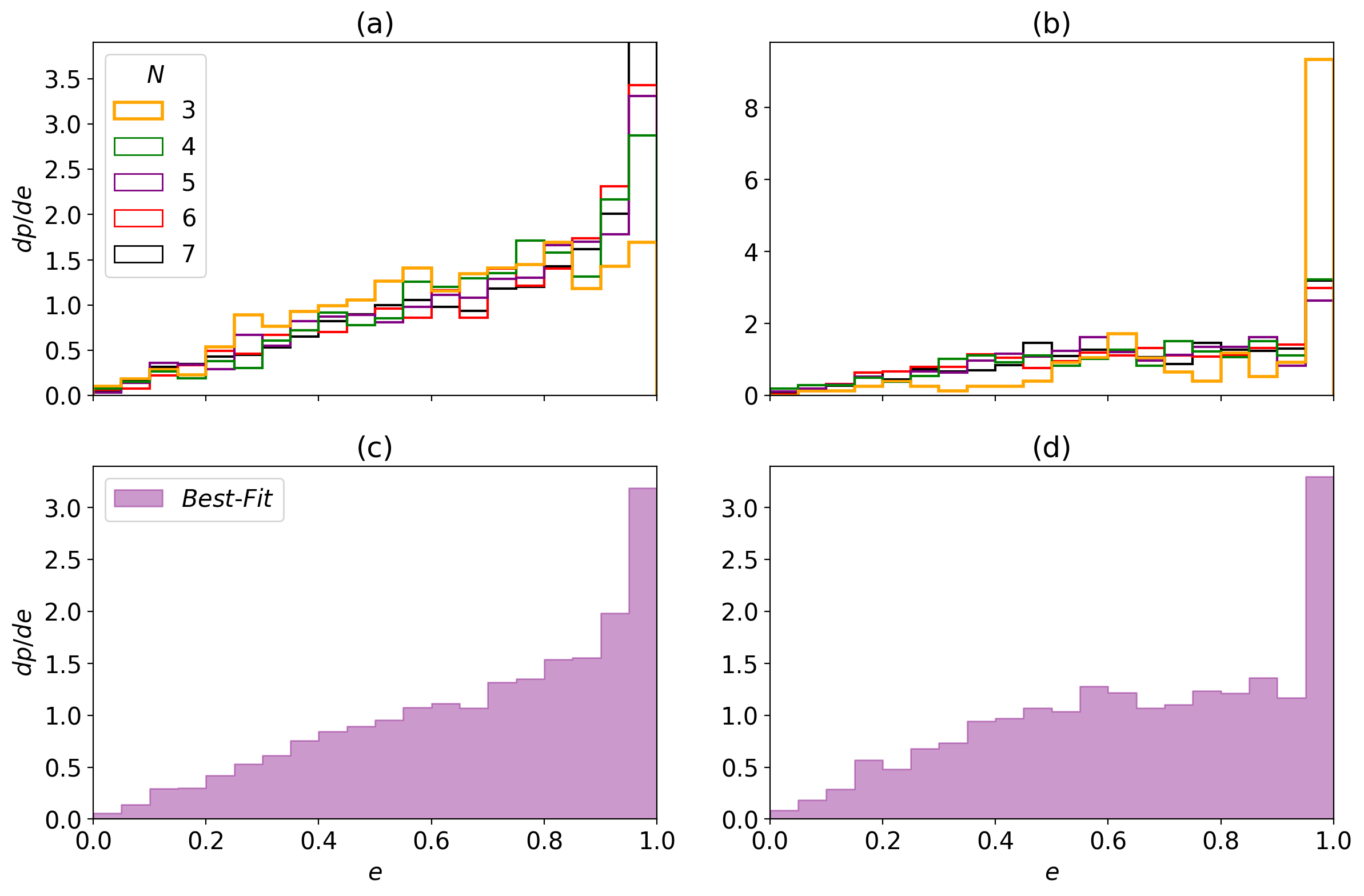}
\hfill    
\caption{The specific probability distributions of eccentricity, $e$. (a) S2 orbits, (b) S1 orbits, for the {\it Fiducial Case} with $N\!=\!3$, $4$, $5$, $6$, $7$. (c) S2 orbits, (d) S1 orbits, for the {\it Best-Fit Case}.}
\label{fig:eComp}
\end{figure*}

Figure \ref{fig:eComp}a shows the specific probability distributions of eccentricity, $e$, for S2 orbits in the {\it Fiducial Case} with $N\!=\!3$, $4$, $5$, $6$, $7$. Figure \ref{fig:eComp}c shows the  distribution of $e$ for S2 orbits in the {\it Best-Fit Case}. In all cases, the distribution of eccentricities is very close to thermal, i.e. $dp/de\!\simeq\!2e$. The only significant departure from a thermal distribution is an excess of very eccentric orbits ($e\!\simeq\!1$), which is evident for all cases except the {\it Fiducial Case} with $N\!=\!3$. For S1 orbits, this population is likely to be both highly unstable to external perturbations, and difficult to observe; this issue is discussed further in Section \ref{sec:highe}.

The high-$e$ excess increases with $N$. Some of this high-$e$ excess may be due to stellar exchanges between nested S2 and S1 orbits, or to tertiary members exciting S2 orbits into higher eccentricities via ZLK cycles. The tertiary may then either remain bound to the excited S2 system, or be ejected. The increase in the high-$e$ excess with increasing $N$ supports these hypotheses, as the opportunities for orbital exchanges, and the number of systems capable of ZLK cycles, both increase with $N$. 

Figures \ref{fig:eComp}b and \ref{fig:eComp}d show the specific probability distributions of $e$ for S1 orbits for, respectively, the {\it Fiducial Case} with $N\!=\!3$, $4$, $5$, $6$, $7$, and the {\it Best-Fit Case}. Again the distributions are approximately thermal, albeit with low counts for the low-$N$ cases, and again there is an excess of high eccentricity orbits ($e\!>\!0.95$).

\subsection{Dynamical Biasing}

\begin{figure}
\centering
\includegraphics[width=0.48\textwidth]{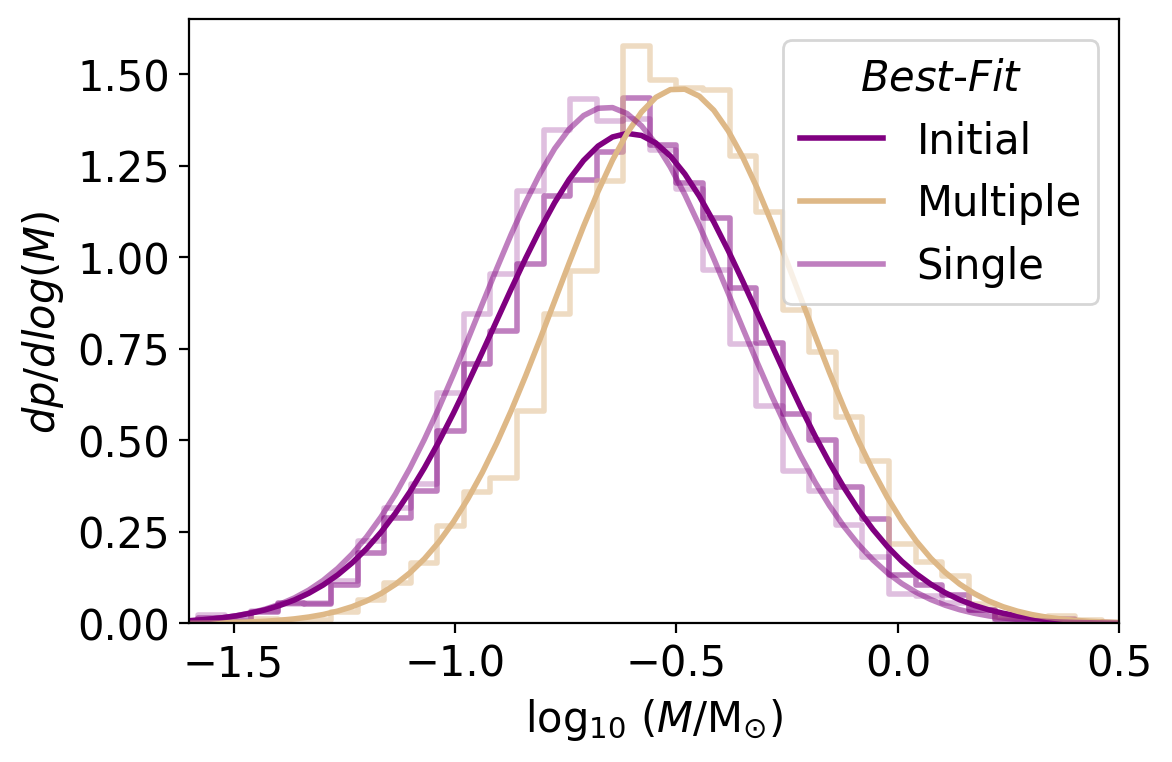}
\hfill    
\caption{The mass distributions for all stars, for those that end up in multiples, and for those that end up single, in the {\it Best-Fit Case}. Means and standard deviations are given in Table \ref{tab:mass}.}
\label{fig:MDist5}
\end{figure}

\begin{table}
\caption{The mean and standard deviation of log-normal fits to the mass distributions for stars that end up in multiple systems and stars that end up single. Values are given for different $N$ in the {\it Fiducial Case} and for the {\it Best-Fit}  $N$-distribution. $\ell\!=\!\log_{_{10}}\!(M/M_{_\odot}$).}
\vspace{-0.3cm}
\begin{center}
\begin{tabular}{rcccccc}\hline
\multicolumn{6}{l}{\it Fiducial} & {\it Best-Fit} \\
with$\;\,N=$ & 3 & 4 & 5 & 6 & 7 & \\\hline
{\sc Initial} & & & & & & \\
$\mu_\ell$ & -0.60 & -0.60 & -0.60 & -0.60 & -0.60 & -0.60 \\
$\sigma_\ell$ & 0.30 & 0.30 & 0.30 & 0.30 & 0.30 & 0.30 \\\hline
{\sc Multiple} & & & & & & \\
$\mu_\ell$ & -0.52 & -0.50 & -0.48 & -0.44 & -0.46 & -0.49 \\
$\sigma_\ell$ & 0.28 & 0.28 & 0.27 & 0.28 & 0.28 & 0.27 \\\hline
{\sc Single} & & & & & & \\
$\mu_\ell$ & -0.71 & -0.68 & -0.65 & -0.62 & -0.59 & -0.66 \\
$\sigma_\ell$ & 0.27 & 0.28 & 0.28 & 0.29 & 0.31 & 0.28 \\\hline
\end{tabular}
\end{center}
\label{tab:mass}
\end{table}

\begin{figure}
\centering
\includegraphics[width=0.48\textwidth]{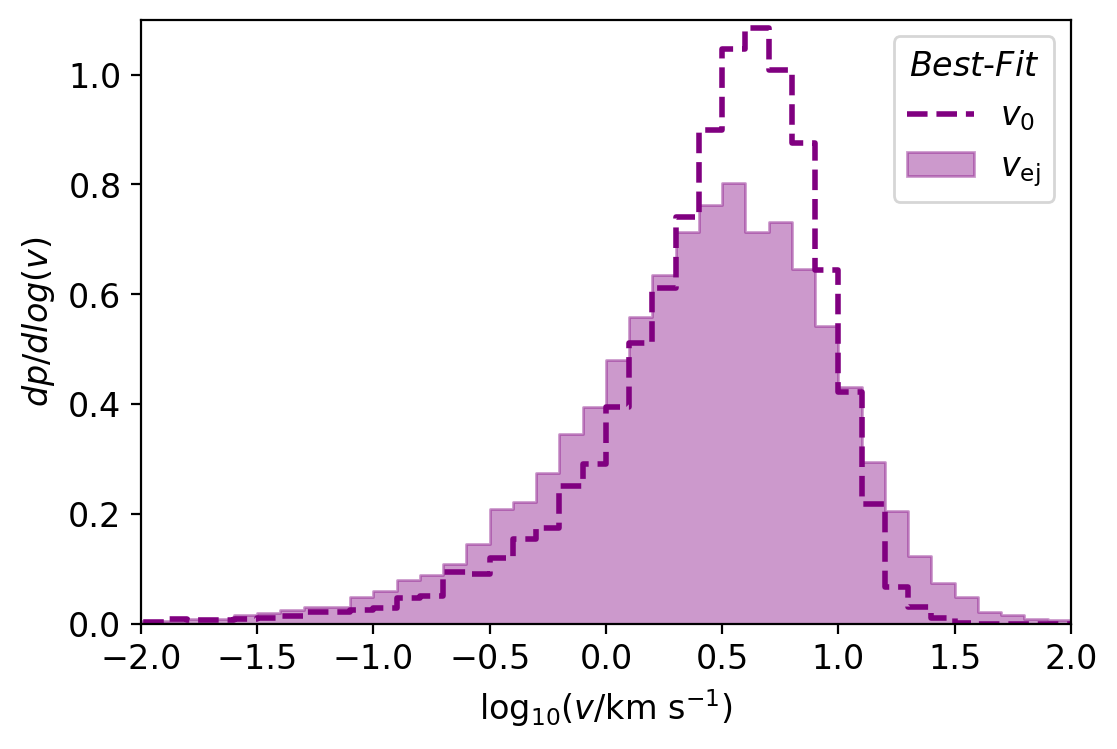}
\hfill    
\caption{The specific probability distributions of initial velocity ($v_0$) and ejection velocity ($v_{\rm{ej}}$) for all stars in the {\it Best-Fit Case}.}
\label{fig:v_ej}
\end{figure}

\begin{figure}
\centering
\includegraphics[width=0.48\textwidth]{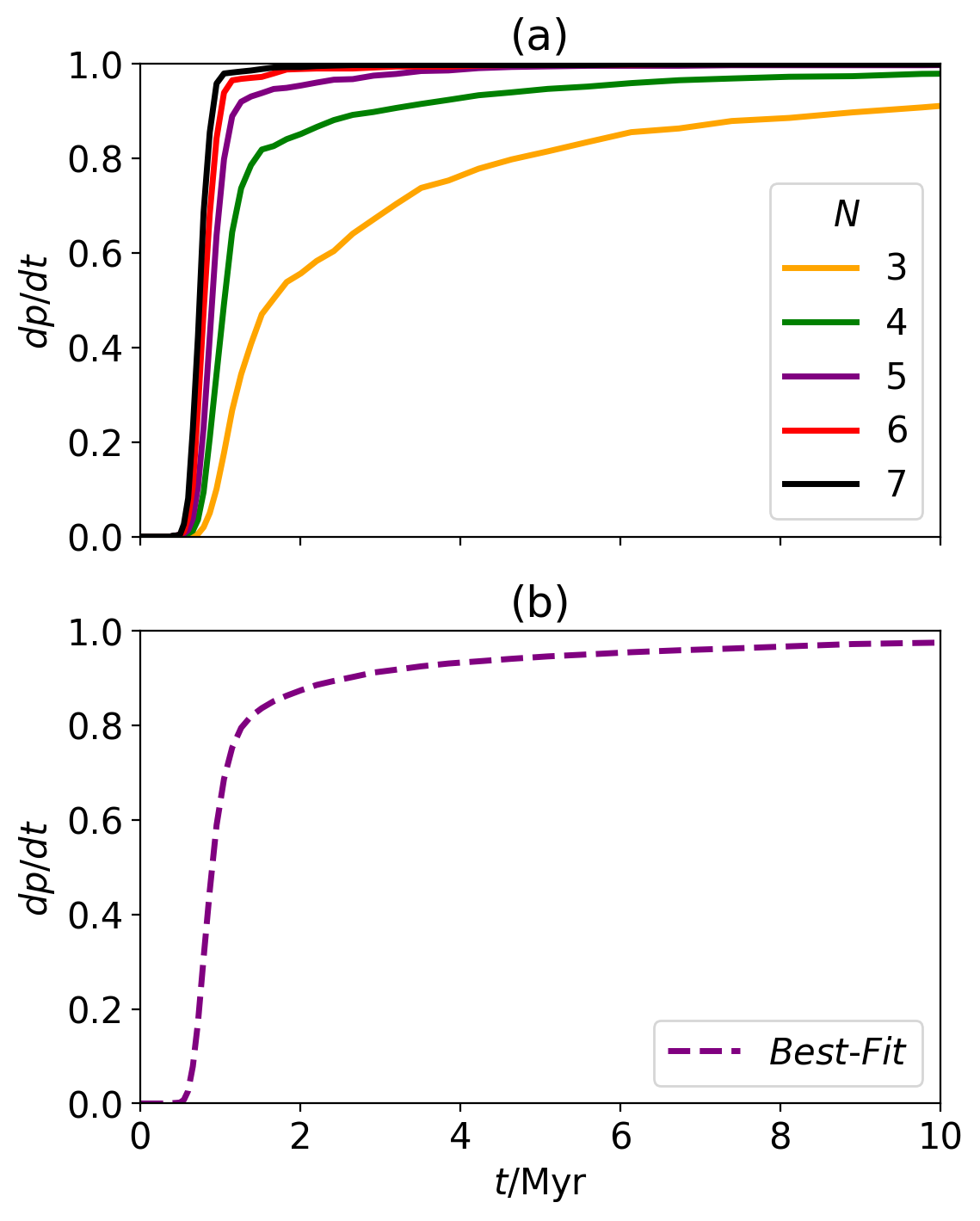}
\hfill    
\caption{The cumulative probability distributions of the first ejection as a function of the elapsed time. (a) The {\it Fiducial Case} with $N\!=\!3$, $4$, $5$, $6$, $7$. (b) The {\it Best-Fit Case}.}
\label{fig:ejN}
\end{figure}

Table \ref{tab:mass} gives the means and standard deviations of log-normal fits to the mass distributions for all stars, for stars that end up in multiples, and for stars that end up single, in the {\it Fiducial Case} with $N\!=\!3$, $4$, $5$, $6$, $7$, and in the {\it Best-Fit Case}. Due to {\it Dynamical Biasing} \citep{1993MNRAS.262..800M}, the stars in multiple systems are in all cases on average more massive than the singles, typically by $\sim\!50\%$. Figure \ref{fig:MDist5} illustrates this for the {\it Best-Fit Case}; the distributions for the Fiducial cases look very similar.

\subsection{Ejection Velocities} \label{sec:esc}

All stars start off as members of virialised subclusters. Therefore at the end any single star must have been ejected fast enough to become unbound from all the other stars in its birth subcluster.

Fig \ref{fig:v_ej} compares the Maxwellian distribution of initial stellar velocities, $v_0$, with the velocities of single stars at the end, $v_{\rm ej}$, for the {\it Best-Fit Case}. The ejection velocity distribution is slightly broadened compared with the initial velocities. Few stars have velocity outside of the initial velocity range. A similar result is found for the {\it Fiducial Case}.

\subsection{Dissolution Timescale}

Figure \ref{fig:ejN}a shows the cumulative probability distribution of single-star ejections as a function of time, for the {\it Fiducial Case} with $N\!=\!3$, $4$, $5$, $6$, $7$. Ejections occur more rapidly for larger $N$, as shown by \cite{2020MNRAS.494.5298H}, but even for $N\!=\!3$, $50\%$ of ejections occur within the first 66 crossing times.

Figure \ref{fig:ejN}b shows the cumulative probability distribution of single-star ejections as a function of time, for the {\it Best-Fit Case}. $\,87(\pm 5)\%$ of ejections occur in the first $2$ Myr, and $98(\pm 2)\%$ occur by 10 Myr.

\subsection{Number of Multiples per Subcluster}

In most cases, a subcluster in the experiment dissolves into one multiple system of multiplicity m, and $N-m$ single stars. But in some instances the subcluster dissolves into as many as $\lfloor N/2 \rfloor$ multiple systems. For the {\it Best-Fit Case}, $21(\pm 1)\%$ of subclusters produce more than one multiple system. All additional multiple systems produced in this case are binaries.

\section{Discussion}\label{discussion}

In this section we discuss some of the caveats and limitations that should be borne in mind when evaluating the results presented above.

\subsection{Dynamical Biasing and Ejections}

If, as we find in AW24, a prestellar core typically produces $N\!=\!4$ or $5$ stars, and these stars have a relatively small range of masses ($\sigell\!\sim\!0.3$), the concept of {\it Dynamical Biasing} is slightly more nuanced, in the sense that even if an S2 orbit does not involve the two {\it most} massive stars in the subcluster, it will still -- more often than not -- involve two of the {\it more} massive stars.

This in turn means that the single stars that are ejected will tend to be of lower mass, and therefore harder to detect in observations than the ones that end up in multiple systems.

\subsection{The Core Potential and Subcluster Dispersal}

It is reasonable that we have only taken account of the gravitational field of the stars in the subcluster. The gas between the stars in the subcluster will only dominate the gravitational field in the region where the stellar dynamics is modelled ($R<R_{\rm o}$) if its mean density is very high,
\begin{eqnarray}
{\bar n}_{_{\rm H_2}}\!&\!\!\gtrsim\!\!&\!2\times 10^9\,{\rm cm^{-3}}\left(\!\frac{N}{5}\!\right)\left(\!\frac{\bar{M}}{0.25\,{\rm M}_{_\odot}}\!\right)\left(\!\frac{R_{\rm o}}{500\,{\rm AU}}\!\right)^{-3}\!.\hspace{0.3cm}
\end{eqnarray}
This is unlikely. However, residual gas from the prestellar core at radii larger than $R_{\rm o}$ will slow down dispersal of the stars formed, unless it has been very widely dispersed. Therefore the ejection velocities we compute should strictly be seen as upper limits.

\subsection{Disc Fragmentation and Primordial Multiples}

The stars in our numerical experiments start with random positions and velocities. Therefore primordial binaries in close, tight orbits will be extremely rare at the outset. However, in nature stars forming by dynamical fragmentation of a prestellar core will be attended by circumstellar accretion discs, and these discs may become sufficiently massive and cold to fragment, producing close companions (semi-major axes $a\lesssim 50\,{\rm AU}$) on low-eccentricity orbits. Indeed, we believe that disc fragmentation is a critical channel for producing close systems \citep[e.g.][]{2006A&A...458..817W}. \cite{2023A&A...674A.196K} have also shown that the gas present around forming stars can trigger their inspiral and create close binaries. These close systems would be more difficult to disrupt, increasing the proportion of higher-order multiples. Therefore the experiments reported here are not expected to reproduce the statistics of the closest observed orbits.

\subsection{Binary Twins}

A significant fraction of observed binary stars have mass ratio close to unity. In other words the component stars have very similar masses. This preference is too extreme to be the result of dynamical biasing, and is normally attributed to hydrodynamical effects, which are not captured in the pure N-body numerical experiments reported here. The standard explanation \citep[][their Section 5.2]{1995MNRAS.277..727W} is that a binary forms, possibly by disc fragmentation, but there is then ongoing accretion onto the binary. The specific angular momentum of the inflowing material increases with time, and therefore the lower-mass component, which has higher specific angular momentum (by a factor $q_{_1}^{-2}$) and is therefore on a wider orbit, is better able to accrete this material. Consequently it grows towards the same mass as the higher-mass component.

\subsection{Interacting Discs and Tidal Circularisation}

An attendant circumstellar accretion disc may also interact with another star and its disc. This will dissipate kinetic energy, and may thereby make the two stars bound, or more tightly bound if they are already bound \citep[e.g.][]{1995MNRAS.275..671M}. This mechanism is missing from our pure N-body experiments, and would help to produce closer orbits.

In addition, stars on extremely close orbits ($a\lesssim 0.1\,{\rm AU}$) will experience strong internal tidal interactions, and this will drive them into low-eccentricity orbits. However, such close orbits are not produced here -- and tidal interactions would not be captured, even if they were, since the stars here are point masses.

Binary orbits may also be circularised when the component stars accrete from a circumbinary disc. Again this is not included in our numerical experiments.

\subsection{Time-scales}

The {\it Orbital Statistics} presented in this paper represent subclusters that have been evolved for 1000 crossing times. With the scalings we have adopted, one crossing time is $\sim\!0.077\,{\rm Myr}\,N^{-1/2}$, which for the {\it Best-Fit Case} with $\mu_N\!=\!4.8$ gives $\sim\!0.035\,{\rm Myr}$. {\it Orbital Statistics} are only considered robust if they are reproduced in two successive MMOs (see Section \ref{SEC:MMO}). Thus the first {\it Orbital Statistics} are collated after 66 crossing times, i.e. at $\sim\!2\,{\rm Myr}$.

For the {\it Fiducial Cases} with large $N\!\geq\!4$, and for the {\it Best-Fit Case}, Figure \ref{fig:ejN} indicates that the {\it Orbital Statistics} are closing in on their asymptotic values by $t\!\lesssim\!3\,{\rm Myr}$, whereas for $N\!=\!3$, the {\it Orbital Statistics} only approach their asymptotic values at $t\!\gtrsim\!15\,{\rm Myr}$. This may constitute a new and -- at least in principle -- distinctive constraint on the size of the region in which the protostellar fragments in a core become an ensemble of virialised protostars, like the ones we have modelled. If we exploit the fact that the experiments are scale-free, i.e. we can adjust the mass- and length-scales as described in Section \ref{SEC:ICs}, then the radius is given by
\begin{eqnarray}
R_{\rm o}\!\!&\!\!\gtrsim\!\!&\!\!250\,{\rm AU}\left(\frac{N}{5}\right)^{2/3}\!\left(\frac{\bar{M}}{0.25{\rm M}_\odot}\right)^{1/3}\!\left(\frac{t_{\rm disp}}{\rm Myr}\right)^{2/3}\!,\hspace{0.3cm}N\!>\!3,\hspace{0.7cm}\\
R_{\rm o}\!\!&\!\!\lesssim\!\!&\!\!60\,{\rm AU}\left(\frac{\bar{M}}{0.25{\rm M}_\odot}\right)^{1/3}\!\left(\frac{t_{\rm disp}}{\rm Myr}\right)^{2/3}\!,\hspace{1.50cm}N\!=\!3,
\end{eqnarray}
where $N$ is the number of stars in the subcluster, $\bar{M}$ is the mean stellar mass, and $t_{\rm disp}$ is the timescale on which the subcluster disperses.

\subsection{External Perturbations} \label{sec:highe}

Even before a subcluster disperses due to internal interactions, it may be subject to external perturbations, due to other massive structures in the vicinity, i.e. other subclusters and gas clumps. These perturbations will sometimes unbind the outer members of the subcluster and will destroy some of the multiples.

Once the subcluster disperses, its stars and multiple systems will interact with other stars and multiple systems formed in nearby subclusters. These interactions will change the architectures of existing multiples, and lead to exchanges of stars between the multiples from different subclusters. In the long term these processes will tend to reduce the overall multiplicity and plurality, by unbinding wider orbits, but at the same time it will harden closer orbits.

The high-eccentricity tertiary orbits noted in Section \ref{SEC:OrbEcc} (see Figure \ref{fig:eComp}) are real in the sense that their orbital parameters are confirmed at successive MMOs. However, successive MMOs are only $\sim\!1\,{\rm Myr}$ apart, and these highly eccentric orbits have periods $\sim\!1\,{\rm Gyr}$. Consequently they will be hard to detect observationally, firstly because they will spend most of their lifetime at distances $\gtrsim\!1\,{\rm pc}$, and secondly because association with a companion will be hard to establish. Therefore they are unlikely to appear in catalogues. More importantly, they have very low binding energy and will be easily disrupted. We find that performing a cut to remove these high-period orbits ($P<10^7$ yr, $a \lesssim 4.5{\ann \times}10^4$ AU) does not affect the distributions of tertiary mass ratio or mutual orbital inclination at a statistically significant level.

\subsection{{\bf Collisions, Mergers, and Tidal Effects}}

Since the stellar particles are treated as point masses, collisions, mergers, and tidal effects between particles are not included in these experiments. Because the subclusters studied here are small ($N\leq7$) with relatively low number density, random chance collisions are extremely unlikely. Situations which might lead to mergers and tides are also rare. In the \textit{Best-Fit} parameter set, for example, stars approach within $0.5$ AU of one another in $\sim 0.2\%$ of realisations. Tidal effects, meanwhile, are only expected to affect solar-mass pairs with separations $\lesssim 0.1$ AU (e.g. \citealt{2019A&A...626A..22J}). While some of these systems have the potential to merge due to tidal dissipation, thereby reducing the proportion of close systems, the effect on the overall statistics is very small.

\subsection{Initial Spatial Distribution}

We invoke a uniform density profile when positioning the stellar particles. Because the number density of particles in each subcluster is so low, the choice of density profile has little effect on the initial distributions.

Our initially rotating subclusters may begin with either a spherical or oblate geometry (see Sec. \ref{SEC:ICs}). We find that these geometries do not produce statistically significant differences in any of the reported metrics.

\section{Conclusions}\label{conclusion}

The initial conditions of a subcluster play an integral role in determining the characteristics of the stellar multiple systems that it spawns. The initial number of stars in the subcluster, $N$, has the greatest effect, influencing periods and separations, dynamical biasing, plurality, mutual orbital inclinations, and ejection timescales. The fraction of kinetic energy in ordered rotation, $\alfROT$, and the degree of mass segregation, $\alfSEG$ also have an effect on some of these statistics.

The distributions of semi-major axis, $a$, for S2 orbits shift to lower values with increasing $N$. This is because S2 orbits are hardened by energy exchange with other stars in the subcluster, often leading to the ejection of these other stars. With higher $N$ there are more `other stars' with which to exchange energy.

In contrast the distributions of $a$ for S1 orbits are essentially independent of $N$.

In the {\it Fiducial Case} the percentage of subclusters that produce an S2 orbit involving the two most massive stars decreases from $72(\pm1.5)\%$ for $N\!=\!3$, to $62(\pm1.6)\%$ for $N\!=\!7$. However, there is still {\it Dynamical Biasing}, i.e. the tendency for more massive stars to be bound in multiples and lower-mass stars to be ejected, when $N$ is large. This is because, when $N$ is large, stars other than the two most massive ones may still be quite massive.

Consequently the single stars tend to have lower than average mass, and to acquire the highest velocities, $\upsilon_{\rm ej}$, relative to the centre of mass of the original subcluster. The distribution of $\upsilon_{\rm ej}$ for these single stars is indistinguishable from the Maxwellian distribution of velocities in the initial subcluster, so they should be classified as `walk-aways', rather than `run-aways'. ~Walk-aways are ejected earlier in subclusters with higher $N$.

S2 orbits have a flat distribution of mass ratios between $q_{_1}\!=\!0.2$ and $q_{_1}\!=\!1.0$. ~S1 orbits have a flat distribution of mass ratios between $q_{_2}\!=\!0.1$ and $q_{_2}\!=\!0.5$ with very few higher values. These distributions do not depend strongly on the initial conditions of the subcluster.

On average, a star's Plurality (i.e. the number of companions that a star has, irrespective of whether it is a primary) increases with its mass. The maximum number of companions increases with $N$, and almost all stars with $M\!\gtrsim\!0.5{\rm M}_{_\odot}$ end up with at least one companion when $N\!\geq\!3$.

Moderate rotation results in triple systems with a distribution of mutual orbital inclinations peaking at $\theta\!\sim\!90^\circ$, in agreement with the observed distributions. For triple systems with $\theta\!>\!90^{\circ}$, the majority ($\sim75\%$) form dynamically in high-inclination orbits without the help of vZKL cycles.

For the {\it Best-Fit Case}, $21(\pm 1)\%$ of subclusters produce more than one multiple system.

When considering {\it Multiplicity Statistics}, i.e. the relative proportions of different multiples (singles, binaries, triples, etc.) AW24 found that subclusters should have a distribution of $N$ values with median $\mu_N\simeq 4.8$; the kinetic energy of the stars should be divided between random isotropic velocities drawn from a Maxwellian distribution and ordered rotation, with comparable amounts in each; and there should be mass segregation. Here we have shown that these properties are also compatible with the observed {\it Orbital Statistics}, i.e. the distributions of semi-major axis, mass ratio, eccentricity and mutual inclination. Moreover, the observed distribution of mutual inclinations also strongly favours  $\mu_N\sim 4.8$ ~(see Figure \ref{fig:Bcomp}).

We have shown that there is a relationship (Equation \ref{EQN:Ro.01}) between the size of the region in which the protostars in a subcluster initially condense out ($R_{\rm o}$), the number of stars in the subcluster ($N$), the mean stellar mass ($\bar{M}$), and the timescale ($t_{\rm disp}$) on which the subcluster disperses. We will explore the consequences of this relationship in a future paper.

\section{Acknowledgements}

Hannah Ambrose is grateful for the support of an STFC doctoral training grant. We thank the referee for a careful and constructive report which helped us to improve the original version of the paper.

\section{Data Availability}

The data underlying this article will be shared on reasonable request
to the corresponding author.

\bibliographystyle{mnras}
\bibliography{bibl.bib}

\end{document}